# Simulation of Metal/Oxide Interface Mobility: Effects of Mechanical Stresses on Geometrical Singularities

V. Optasanu[1,a], L. Raceanu[1,b] and T. Montesin[1,c]

[1]Université de Bourgogne, UMR 5209 CNRS, 9 Avenue A. Savary, Dijon, France
[a]virgil.optasanu@u-bourgogne.fr, [b]laura.raceanu@u-bourgogne.fr,
[c]tony.montesin@u-bourgogne.fr

**Keywords:** stress diffusion coupling, oxidation under mechanical stress, waved metal-oxide interface

**Abstract.** During the last decade, an increasing importance has been given to the feedback of mechanical stresses on the chemical diffusion and, further, on corrosion. Many works point the active role of stresses on the material ageing especially on their negative consequences leading to the damaging of structures. Based on a theoretical study and using numerical tools and experimental results our previous works [1,2] on stress/diffusion coupling, highlight the strong influence of stress field on the diffusion process. The aim of the present paper is to describe the influence of some particular morphologies of the metal/oxide interface on both diffusion and oxidation process. The oxidation is assumed to be driven by a mass conservation law (Stefan's law) while the diffusion coefficient of oxygen in metal is locally influenced by the stress field. The stability of a waved-shape interface is studied in both cases: simple diffusion and coupled stress/diffusion process. In this purpose we have developed an original numerical model using a virtual metal/oxide interface of a mono-material with oxygen concentration-dependent parameters, which allows to operate easily with any shape of interface and to use simple finite element meshes. Furthermore, in order to underline in a more obvious way the consequences of mechanical stress on the diffusion process, a particular geometry is studied.

**Introduction**

In order to manage correctly the design of the industrial components and the prediction of their time life, the simulation of the corrosion process presents a particular interest. Whilst the process can be rather easy to describe in simple mechanical conditions, sometimes, the mechanical stress configuration, with localized high stress values and especially high stress concentrations gives a significantly different behavior of materials. The simulation of the oxidation, based on phase field method, already exists in the literature but for the moment no effect of mechanical stress is considered [3].

Stress can be generated by multiples factors, the most obvious one being of course the external loadings. Self induced stress is also an important factor that changes locally the mechanical state and that is supposed to be one of the reason of particular-shaped interfaces between oxide and substrate. One of that particular shape is the waved interface between oxide and metal.

The oxidation process generally occurs with volume changing of the corrosion product compared to the original volume of metal involved in the chemical reaction. The change of the volume is characterized by the Pilling-Bedworth ratio [4] representing the ratio between the corrosion product volume and the original metal volume before oxidation. This parameter can be sensibly higher than 1, which can generate strong compressive mechanical stress in the oxide layer. Initially, a small perturbation can be developed, producing an original defect in the shape of the oxide/substrate interface. After a substantial oxidation time the morphology of the metal/oxide interface can be completely different from the original one. Generally, the shape of the interface is waved.

Explanations were proposed [5,6] based on minimization of mechanical energy of the oxide/metal system which comes with the expansion of the substrate during the transformation into oxide. Actually, the elastic energy stored within a waved layer is lower than in a planar shape layer because of the high values of the compressive stress and strain that have to be developed in that last configuration. Relaxation and creep occurring in the metal near the interface is supposed to be favorable to the bending or buckling of the oxide layer.

During the last decade, the theory of interdependence between chemical diffusion and mechanical stresses has been established using thermodynamic tools for anionic oxidation process. Mechanical stress influences on the effective diffusion coefficient and can be summarized by [7,8]:

$$D_{eff} = D_0 \left[1 - \frac{M_0 \eta_{ij} c}{RT}\left(\frac{\partial \sigma_{ij}}{\partial c} + \frac{\sigma_{ij}}{c}\right)\right] \quad (1)$$

where $D_0$ is the nominal diffusion coefficient, $M_0$ is the molar mass of the metal, $\eta_{ij}$ is the chemical expansion coefficient, $R$ is the universal constant, $T$ is the temperature, $c$ is the oxygen concentration and $\sigma_{ij}$ is the mechanical stress.

In our recent works [9] on the oxidation of $UO_2$ in $U_3O_7$ we highlighted that in particular geometrical conditions (represented by multi-domains with different crystallographic orientations of a constant-thickness $U_3O_7$ layer on a mono-crystalline $UO_2$ substrate) the mechanical stress can generate strong local non-homogeneities in the oxygen concentration field as seen in Fig. 1. Thus, surfaces with the same value of concentration shows undulations: some regions are enriched in oxygen while some others are impoverished which may imply that some regions will be oxidized before some others neighboring regions. If stress can generate differences in the rate of oxidation, can it be a factor of stabilization or not of a wave-shaped metal/oxide interface?

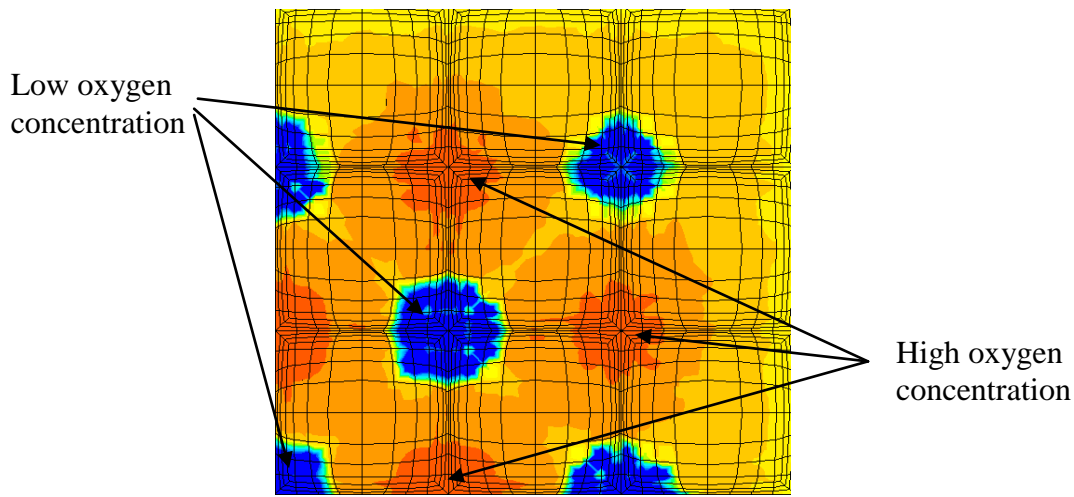

Fig. 1. Concentration field near a planar metal/oxide interface between UO2 substrate and a multi-domain U3O7 layer [9]

The paper will briefly remind the main points of the diffusion under mechanical stress model presented in previous works. Based on Stefan's law with diffusion coefficient depending on stress gradient or stress value, a simple oxidation model with mobile interface between metal and oxide will be presented in this paper.

**Oxidation model**

The simulations have been made using a simple diffusion finite element code (thermic module of CAST3M FEM code) by considering a simple management of the interface between metal and oxide based on the oxygen concentration value. The Fig. 2 shows an example of oxygen concentration profile with respect to the depth. The oxygen concentration has significantly higher values in the oxide than in the metal. It is difficult to manage matter flow with discontinuities in the value of the concentration with classic finite element methods.

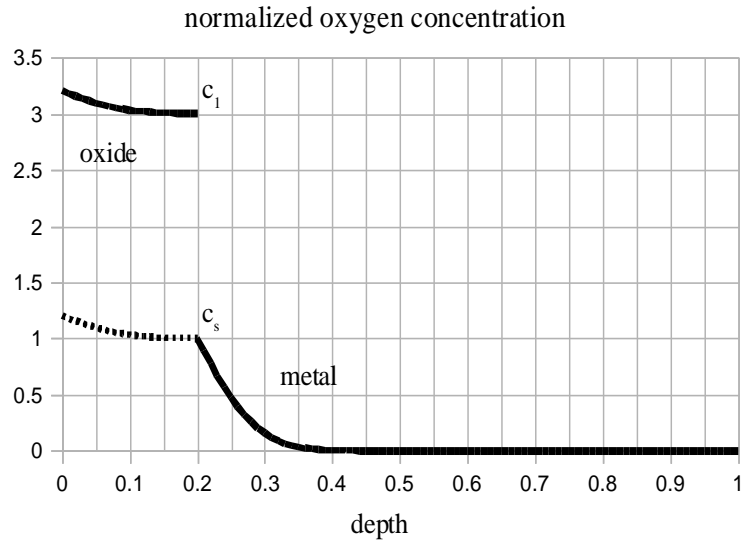

Fig 2. Normalized oxygen concentration.

In order to manage complex shapes of interfaces between metal and oxide the frontier is not a physical one but virtual frontier and it is given by the value of the concentration. Jumps in material parameter values are used in order to simulate the two different materials: metal and oxide. In order to avoid the gap between maximal oxygen concentration value in metal ($c_s$) and respectively its minimal value in oxide ($c_1$) it is possible to suppress that gap (see dashed curve on Fig. 2) and hence to have a continuous curve of oxygen concentration and to use equivalent thermal phase changing operators to simulate that gap in diffusional phenomena. Furthermore, if the concentration is less or equal to $c_s=1$ then the material is metal; while if concentration is superior to $c_s$ the material is oxide. The simulation is mono-material, with concentration-dependent parameters. The parameters reach metal values for concentrations less or equal to $c_s=1$ those parameters reach metal values while when it is superior to $c_s$ they reach oxide values. The jumps in the values of material parameters can be done using a smooth function.

The mobility of the interface is given by a simple Stefan's mass conservation law. If coupling effects between mechanical stress and chemical diffusion are not taken into account, the mass gain obtained with this model is perfectly parabolic.

**Oxidation model**

Let consider an initially waved interface between metal and oxide in a 2D configuration. A rectangular sample with an initial layer of oxide with sinusoidal variable thickness continue to oxidize. Diffusion, geometric and elastic normalized parameters are given in Table 1. Lateral mechanical and chemical symmetries are taken into account.

|  | oxide | metal |
|---|---|---|
| Diffusion Coefficient | Dox / Dmet = 20 | |
| Geometry | Average thickness / wavelength = 1<br>waves amplitude / average thickness = 0.15 | Semi-infinite |
| Expansion coefficient | Pilling-Bedworth ratio = 1.005 | Normalized chemical expansion coef.= 1.6e-2 |
| Elastic coefficients | Young modulus of oxide / Young modulus of metal = 1 | |

Table 1. Simulation parameters

If no feedback of stress on the diffusion coefficient is considered, an initially wave-shaped interface becomes planar after some time as shown in Fig. 3.

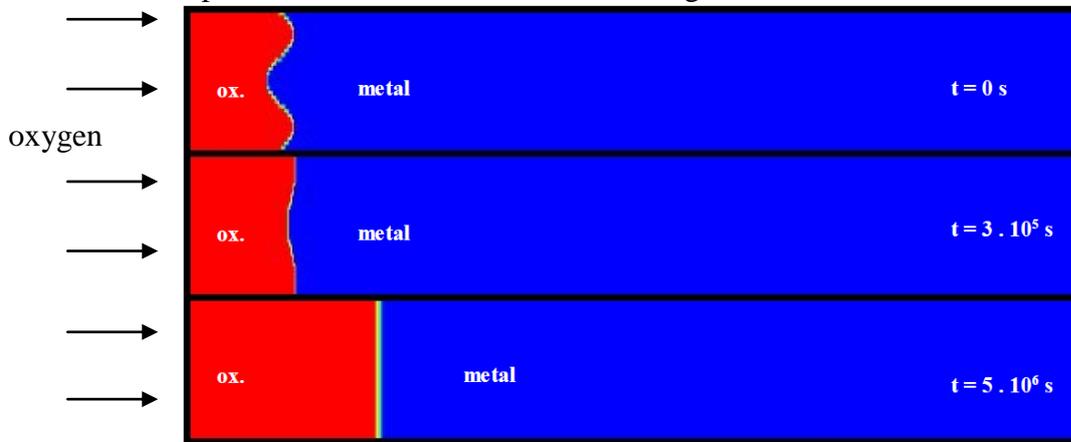

Fig. 3. Simulation of oxidation of an initial waved-shaped metal/oxide frontier.

Such a configuration of the interface implies particular stress state near the interface. Fig. 4 shows the stress in horizontal and vertical directions computed for $Zr / ZrO_2$ system. The stress state considers the Pilling-Bedworth expansion of the oxide and the chemical expansion of the metal enriched with oxygen. Thus, the stress type and intensity are presented in Table 2. In comparison with the work published by Parise et al. [10] we can remark some differences. In locations noted a, b and d, stresses have the same nature, but in location c, in the horizontal direction, we found moderate compressive stress while Parise found moderate tensile stress. This can be explained by the fact that we consider the chemical expansion of the metal enriched with oxygen (while Parise doesn't) and by the lateral symmetries that simulate an infinitely large domain with expanding materials. In a short domain, oxide can expand in horizontal direction and it could involve tensile stress in metal.

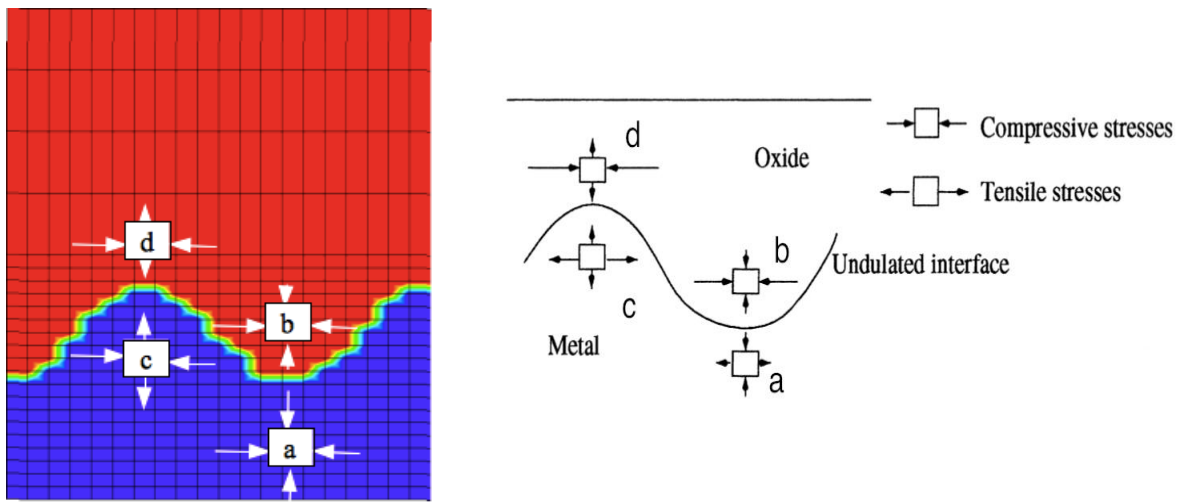

Fig. 4. Stress in the oxide layer and in the substrate. Left - present work, right - Parise [10]

| location | Horizontal direction | Vertical direction |
|---|---|---|
| a | Moderate compression | Moderate compression |
| b | Very strong compression | Low compression |
| c | Moderate compression | Moderate traction |
| d | Strong compression | Low traction |

Table 2. Stress type and intensities

In order to highlight the influence of each term of Eq. (1) we compute separately, first the stress gradient term and then the term containing the stress.

By taking into account the first coupling term in Eq. (1), those containing the stress gradient, after 1500s of diffusion we can remark slight differences between the shapes of the interface in the case of the coupled stress diffusion (Fig. 5.c) and simple diffusion (Fig. 5b). The stress gradient tends to accelerate the oxidation. In the case of no stress coupling we can remark that the interface is less wavy than initially. The undulation is more persistent in the case of a diffusion coefficient depending on the stress gradient, which means that stresses have a stabilization effect on the undulations. The general shape of undulations changes but the amplitude vanishes slower.

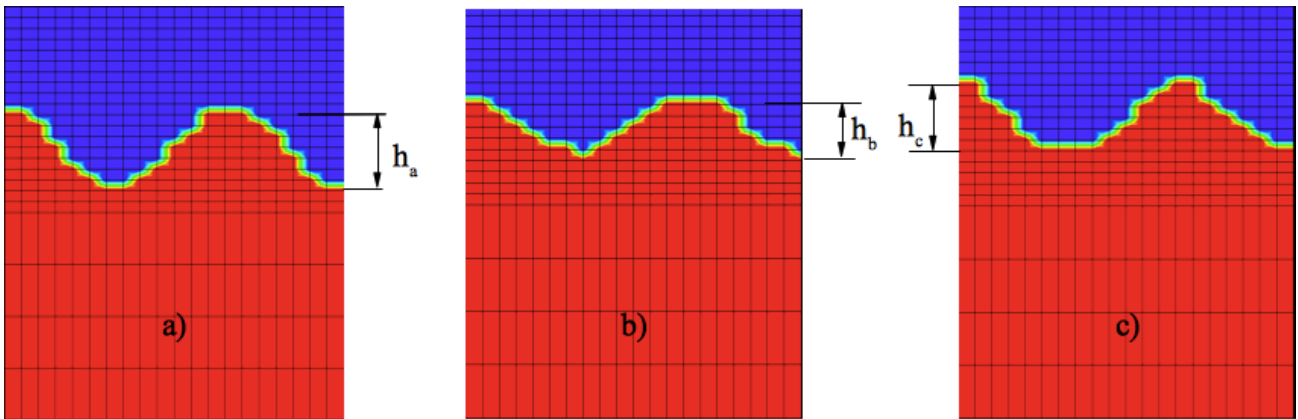

Fig. 5. Simulations of oxidation: a) initial configuration, b) oxidation interface after t = 1500 s without taking into account the stresses, c) oxidation interface after t = 1500 s with diffusion coefficient depending on the stress gradient.

Unfortunately, as the computations are very time-consuming when the stress gradient is taken into account (small time steps) a general conclusion about the final shape of the interface after several hours of oxidation cannot be reached.

If, in Eq. 1, the stress term is taken into account and if we neglect the stress gradient term, effects of stress are not on the amplitude of undulations but on the interface velocity. Thus, stress term seems to decrease the rate of oxidation. Computations are much less time-consuming than those with stress gradient.

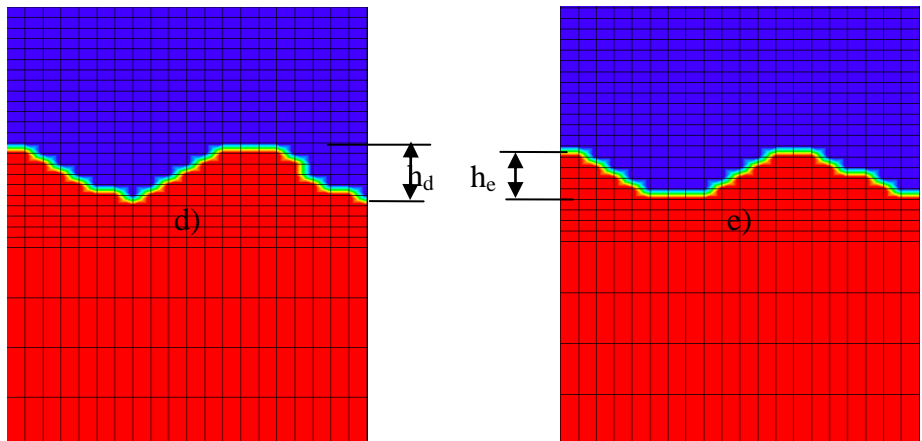

Fig. 6. Simulations of oxidation: d) oxidation interface after t = 1500 s without taking into account the stresses, e) oxidation interface after t = 1500 s with diffusion coefficient depending on the stress.

## Conclusion

The model presented in this paper is based on a simple management of the interface between metal and oxide using the oxygen concentration value. The frontier between oxide and metal is virtual and given by a threshold value which is the saturation concentration of the oxygen in the metal. All parameters are a function of the oxygen concentration and present jumps in order to describe in the same time the oxide and the metal. An initially waved interface becomes planar after few hours of oxidation if no influence of mechanical stresses on the diffusion is taken into account. If gradient of the stress is considered, we shown that the oxidation is more rapid and that the waves are more persistent. If the stress term only (and not the gradient stress term) in Eq. 1 is considered we can notice a decrease of the average rate of oxidation.

The next step to be done will be to take both terms simultaneously in our computations. An optimization of the model have to be done in order to increase the rapidity of the simulations and reach substantial time of oxidation in order to reach firm conclusions about the persistence of the undulations. At the same time a parametric study will be done in order to highlight the relations between diffusion parameters and geometrical shape of the interface which could leads to an increase of the undulations amplitude. If the results of this future works will be inconclusive this tools will be used to study the growing of the $ZrO_2$ layer in nuclear clads systems, which presents such an undulations at the metal/oxide interface.

**References**

[1] N. Creton, V. Optasanu, T. Montesin, S. Garruchet and L. Desgranges, Defect and Diffusion Foum Vol. 447 (2009), p. 289

[2] N. Creton, V. Optasanu, S. Garruchet, L. Raceanu, T. Montesin, L. Desgranges and S. Dejardin: Defect and Diffusion Foum. Vol. 519 (2010), p. 297

[3] K. Ammar, B. Appolaire, G. Cailletaud, F. Feyel and S. Forest: Computational Material Science Vol. 45-3, (2009), p. 800

[4] N.B. Pilling and R.E. Bedworth, J. Inst. Met. Vol. 29 (1923), p. 529

[5] R.J. ASaro, W.A. Tiller, Metall. Trans. Vol. 3 (1972), p.1789.

[6] M.A. Grinfeld': Sov. Phys. Dokl. Vol. 11 (1986), p 31.

[7] J. Favergeon, T. Montesin and G. Bertrand: Oxidation of Metals, Vol. 64-3/4 (2005), p. 253

[8] S. Garruchet, A. Hasnaoui, O. Politano, T. Montesin, J.M. Salazar, G. Bertrand and H. Sabar: Defect and Diffusion Forum Vols 237-240 (2005), p145

[9] L. Desgranges, H. Palancher, M. Gamaléri, J.S. Micha, V. Optasanu, L. Raceanu, T. Montesin and N. Creton: Journal of Nuclear Materials Vol. 402 (2010), p.167

[10] M. Parise, O. Sicardy, G. Cailletaud: J. Nuc. Mat., Journal of Nuclear Materials Vol 256 (1998), p.35